\documentclass[journal]{IEEEtran}
\usepackage[T1]{fontenc}
\usepackage{amsmath}
\usepackage{amsfonts} 
\usepackage{pgfplots}
\usepackage{cite}
\pgfplotsset{compat=1.5}
\usepackage{pgfplotstable}
\usepgfplotslibrary{statistics}
\pgfplotsset{grid style={dotted,gray}}
\usepackage{tikz}
\usepackage{xcolor}
\usepackage{caption}
\usepackage{soul}
\newtheorem{prop}{\textbf{Proposition}}
\usepackage{titlesec}

\titlespacing*{\subsection}{0pt}{2pt}{2pt}
\titlespacing*{\section}{0pt}{4pt}{4pt}

\begin{document}

\title{Structured Superposition of Autoencoders for\\ UEP Codes at Intermediate Blocklengths} 

\author{Vukan Ninkovic,~\IEEEmembership{Member,~IEEE,}
        Dejan Vukobratovic,~\IEEEmembership{Senior Member,~IEEE}
\thanks{V. Ninkovic is with the University of Novi Sad, Serbia, and  the Institute for Artificial Intelligence
 Research and Development of Serbia (e-mail: ninkovic@uns.ac.rs);
 D. Vukobratovic is with the University of Novi Sad,
 Serbia, (email: dejanv@uns.ac.rs).}
\thanks{This work is funded by Serbian Ministry of Science, Technological Development and Innovation, through the Science and Technological Cooperation program Serbia-China, project No 00101957 2025 13440 003 000 620 021.}
}

\markboth{Journal of \LaTeX\ Class Files,~Vol.~14, No.~8, August~2015}%
{Shell \MakeLowercase{\textit{et al.}}: Bare Demo of IEEEtran.cls for IEEE Journals}


\maketitle

\begin{abstract}
Unequal error protection (UEP) coding that enables differentiated reliability levels within a transmitted message is essential for modern communication systems. Autoencoder (AE)-based code designs have shown promise in the context of learned equal error protection (EEP) coding schemes. However, their application to UEP remains largely unexplored, particularly at intermediate blocklengths, due to the increasing complexity of AE-based models. Inspired by the proven effectiveness of superposition coding and successive interference cancellation (SIC) decoding in conventional UEP schemes, we propose a structured AE-based architecture that extends AE-based UEP codes to substantially larger blocklengths while maintaining efficient training. By structuring encoding and decoding into smaller AE subblocks, our method provides a flexible framework for fine-tuning UEP reliability levels while adapting to diverse system parameters. Numerical results show that the proposed approach improves over established achievability bounds of randomized superposition coding-based UEP schemes with SIC decoding, making the proposed structured AE-based UEP codes a scalable and efficient solution for next-generation networks.

\end{abstract}


\IEEEpeerreviewmaketitle

\section{Introduction}

\IEEEPARstart{A}{utoencoder} (AE)-based communication systems have gained significant attention in recent years \cite{OShea_2017, shin_2024, song_2022}, demonstrating their potential in designing end-to-end learned transmission schemes. However, their application to unequal error protection (UEP) codes \cite{Masnick_1967} remains largely unexplored, particularly at intermediate blocklengths. A major challenge in extending conventional AE-based architectures to higher blocklengths arises from the curse of dimensionality, where larger blocklengths demand models of exponential complexity. As a result, training AEs for intermediate and long blocklengths becomes impractical due to excessive computational demands and convergence difficulties \cite{Larue_2022}.

 UEP coding can be approached at different levels, notably through message-wise and bit-wise protection schemes \cite{Borade_2009}. While the former assigns different reliability levels to entire messages based on priority, the latter, examined in this work, focuses on protecting specific segments within a binary message according to their significance \cite{Borade_2009, Yao_2024, sheldon_2024}. This property is particularly valuable in multimedia communications \cite{Sejdinovic_2009} or 5G cellular networks, where diverse services such as ultra-reliable low-latency communication (URLLC) and enhanced mobile broadband (eMBB) require distinct reliability levels \cite{Yao_2024}.
Despite their practical significance, the design and analysis of bit-wise UEP codes, including the derivation of achievable and converse bounds on error probability regions at finite blocklengths, remains a challenge \cite{Borade_2009, Yao_2024, sheldon_2024}.

In this paper, we propose a structured AE-based bit-wise UEP code design for intermediate blocklengths. Our inspiration stems from conventional coding theory, since superposition coding and successive interference cancellation (SIC) decoding are provably effective for bit-wise UEP \cite{sheldon_2024, Karimzadeh_2019}. Motivated by these results and our previous work on UEP and superposition AE-based codes \cite{ninkovic_noma, Ninkovic_2021}, our aim is to develop a learning-based UEP code design that retains advantages of superposition-based codes and SIC decoding while addressing scalability challenges at increased blocklengths.

To achieve this, we introduce a structured UEP encoding/decoding process that organizes the code into a smaller set of constituent AE subblocks. Furthermore, we incorporate a compound loss function, originally introduced in \cite{Ninkovic_2021}, to optimize the trade-off between different reliability classes. We show that the proposed method outperforms  randomized superposition coding-based UEP schemes with SIC decoding \cite{sheldon_2024}, expanding the achievable error probability region. Prior studies have shown that the achievability bound from \cite{sheldon_2024} and the practical approach from \cite{Ninkovic_2021} surpass UEP schemes based on the superposition of random Gaussian codes \cite{Karimzadeh_2019}. Additionally, in contrast to our prior work \cite{ninkovic_noma, Ninkovic_2021}, the proposed structured AE-based design enables practical, scalable UEP coding at intermediate blocklengths, eliminating the complexity bottleneck of single-AE architectures and allowing benchmarking against theoretical bounds. By providing precise control over UEP error probability trade-offs, our structured UEP design enhances flexibility and increases the domain of applicable blocklengths, making AE-based UEP coding a viable solution for practical applications, including emerging 6G semantic communication systems \cite{zhong_2024}.


We use the following conventions for the notation. The capacity of a point--to--point Gaussian channel with signal-to-noise ratio (SNR) $\gamma=E_\mathrm{b}/N_0$ is expressed as $C(\gamma)=\frac{1}{2}\text{ln}(1+\gamma), \gamma\geq0$, where $E_\mathrm{b}$ represents the energy per bit and $N_0$ denotes the noise power spectral density. The Gaussian $Q$-function is defined as $Q(x)=\int_x^{+\infty}\frac{1}{\sqrt{2\pi}}e^{-\frac{t^2}{2}} \mathrm{d}t, x\in\mathbb{R}$. The second-order behavior of multi-user Gaussian channels is characterized by the cross-dispersion function defined as $V(x,y)=\frac{x(2+y)}{2(1+x)(1+y)}, x\leq0\leq y$.  For the point-to-point case, the Gaussian dispersion function simplifies to $V(x)=V(x,x)=\frac{x(2+x)}{2(1+x)^2}, x\geq0$.

\section{Background \&\ System Model}

\subsection{System Model}
\label{sec:Model}
We investigate the conventional point-to-point communication setup, where a message $m$ is selected from a finite set $\mathcal{M} = \{1, 2, \dots, M\}$ and transmitted over a noisy channel. To each message $m \in \mathcal{M}$, we assign a corresponding binary sequence $\boldsymbol{s} = (s_1, s_2, \dots, s_k)$, with $k = \log_2(M)$. The encoding process is a mapping $f: \mathcal{M} \to \mathbb{R}^n$ that transforms the message $m$ into a codeword $\boldsymbol{x} = f(m) = (x_1, x_2, \dots, x_n)$, represented as an $n$-dimensional vector. The code rate is defined as $R=k/n$ [bits per channel use]. Each codeword satisfies a fixed energy constraint, i.e., $\boldsymbol{x} \in \mathcal{X} = \{\boldsymbol{x} \in \mathbb{R}^n:\| \boldsymbol{x} \|_2^2=n\}$. For a given $\boldsymbol{x} \in \mathbb{R}^n$, the channel $\mathcal{W}$ outputs $\boldsymbol{y}=(y_1,y_2,\ldots,y_n) \in \mathbb{R}^n$ as a random vector governed by the probabilistic channel law $p(\boldsymbol{y}|\boldsymbol{x})$. The decoder, defined as $g: \mathbb{R}^n \to \mathcal{M}$, estimates the transmitted message as $\hat{m} = g(\boldsymbol{y})$. The objective is to optimize the encoding-decoding pair $(f, g)$ for the channel $\mathcal{W}$ to minimize the average message error probability:
\begin{align}
\label{messag_error_eq}
P_{\textrm{e}} = \frac{1}{M} \sum_{m \in \mathcal{M}} \mathbb{P}\{\hat{m} \neq m|m\}.   
\end{align}

\textit{1) Autoencoder--Based Representation:} The considered communication system can be represented as an AE, consisting of an encoder, a noisy channel, and a decoder \cite{OShea_2017}.
The encoder implements a mapping $\boldsymbol{x}=f(m)$, where a discrete message $m \in \mathcal{M}$ is first encoded as a one-hot vector $\boldsymbol{u} \in \{0,1\}^M$ (a representation in which only the $m$-th position is set to one, while all other elements remain zero), then processed by a feed-forward neural network, and finally passed through a bottleneck layer of dimension $n$. A normalization layer ensures that the codeword satisfies a fixed energy constraint ($\boldsymbol{x} \in \mathcal{X}$). The codeword $\boldsymbol{x}$ is transmitted through a channel $\mathcal{W}$, modeled as an additive white Gaussian noise (AWGN) channel, where the received signal is given by $\boldsymbol{y} = \boldsymbol{x} + \boldsymbol{z}$, with $\boldsymbol{z} \sim \mathcal{N}(\boldsymbol{0}, \sigma^2)$ representing $n$ independent and identically distributed (i.i.d.) noise samples.


The decoder $\hat{m} = g(\boldsymbol{y})$ is implemented as a feed-forward neural network with a symmetric structure to that of the encoder. The final layer employs a softmax activation function, producing a probability vector $\boldsymbol{b} \in (0,1)^M$, where $\sum_{i=1}^{M} b_i = 1$. The estimated message is determined as $\hat{m} = \arg\max_{i} \{ b_i \}$. The AE is trained end-to-end using stochastic gradient descent (SGD) with the Adam optimizer \cite{adam}, minimizing the cross-entropy loss:
\begin{align}\label{eq2}
\ell(\boldsymbol{u},\boldsymbol{b}) = -\sum_{i=1}^{M} u_i \log{b_i},
\end{align}
which serves as a differentiable surrogate for the message error probability $P_{\textrm{e}}$. The loss is averaged over training batches to jointly optimize the encoder and decoder ($(f,g)$ pair), enabling robust message reconstruction under channel impairments.

\subsection{Bit--Wise Unequal Error Protection Code Design}
\label{bit_wise}
In the context of bit-wise UEP, an alternative representation $\mathcal{S}$ of the message set $\mathcal{M}$ is preferred, where $\mathcal{S}$ consists of binary sequences $\boldsymbol{s}$ (Section~\ref{sec:Model}). The sequence $\boldsymbol{s}$ is partitioned into $C$ distinct submessages, $\boldsymbol{s} = (\boldsymbol{s}_1, \boldsymbol{s}_2, \dots, \boldsymbol{s}_C)$. The length of the $i$-th submessage $\boldsymbol{s}_i$ is $k_i$ bits, while the length of the whole sequence is $k = \sum_{i=1}^C k_i$. Consequently, $|\mathcal{S}| = 2^{k}$ and $|\mathcal{S}_i| = 2^{k_i}$. In this framework, the individual code rates for each class are defined as $R_i=k_i/n, i\in\{1, \dots,C\}$ and $R=\sum_{i=1}^CR_i$.

In the bit-wise UEP approach, submessages are considered to belong to different message classes having different importance, with each message class receiving a different level of error protection. Consequently, the probability of error is defined separately for each message class to reflect its specific protection level. Each submessage $\boldsymbol{s}_i \in \mathcal{S}_i$ is associated with a set $\mathcal{M}_{\boldsymbol{s}_i}$, which includes all messages $m \in \mathcal{M}$ whose $i$-th submessage in $\boldsymbol{s}$ matches $\boldsymbol{s}_i$. Given an encoder-decoder pair $(f, g)$, the per-class error probability is expressed as:

\begin{align}
    P_{\textrm{e}}^{(i)}=\frac{1}{|\mathcal{S}_i|} \sum_{\boldsymbol{s}_i \in \mathcal{S}_i} \mathbb{P}\{\hat{m} \notin \mathcal{M}_{\boldsymbol{s}_i}| m \in \mathcal{M}_{\boldsymbol{s}_i}\},
\end{align}
where the per--class error probabilities are collected in a vector  $\boldsymbol{P}_{\textrm{e}}=(P_{\textrm{e}}^{(1)}, P_{\textrm{e}}^{(2)}, \ldots, P_{\textrm{e}}^{(C)})$. Finally, the formal definition of bit--wise UEP codes is given by triplet $(\{\mathcal{S}_i\}_{i=1}^C,f,g)$.

\textit{1) Autoencoder--Based UEP Code Design:} In \cite{Ninkovic_2021}, we extended the AE-based communication system defined in \cite{OShea_2017} to support the UEP framework.  We introduced a compound loss function that generalizes the cross-entropy function in Eq. \eqref{eq2}. For a message $\boldsymbol{s} = (\boldsymbol{s}_1, \boldsymbol{s}_2, \dots, \boldsymbol{s}_C)$, we define $C$ independent loss functions, where the loss function corresponding to the $j$-th submessage class is given as:
\begin{align}\label{eq4}
\ell(\boldsymbol{u}_{\boldsymbol{s}_j},\boldsymbol{b}) = -\sum_{i=1}^M u_i \log{b_i},
\end{align}
where $\boldsymbol{u}_{\boldsymbol{s}_j} = (u_1, u_2, \dots, u_M)$ is a modified version of the conventional one-hot encoded vector $\boldsymbol{u}$ adapted to submessage $\boldsymbol{s}_j \in \mathcal{S}_j$. Specifically, the $m$-th entry of $\boldsymbol{u}_{\boldsymbol{s}_j}$ is set to one if the binary sequence representation $\boldsymbol{s}$ of message $m$ contains the $j$-th submessage $\boldsymbol{s}_j$. The set $\mathcal{U} = \{\boldsymbol{u}_{\boldsymbol{s}_1}, \boldsymbol{u}_{\boldsymbol{s}_2}, \ldots, \boldsymbol{u}_{\boldsymbol{s}_C}\}$ consists of the modified one-hot vectors for all $C$ submessages in $\boldsymbol{s}$. 
Under this formulation, the compound loss function for the bit-wise UEP case is expressed as follows:
\begin{align}\label{eq3}
    \ell(\mathcal{U},\boldsymbol{b})=\sum_{j=1}^C \lambda_j \ell(\boldsymbol{u}_{\boldsymbol{s}_j},\boldsymbol{b}).
\end{align}
The vector $\pmb{\lambda} = (\lambda_1, \lambda_2, \ldots, \lambda_C)$ collects the weights assigned to each of the message classes, where $\lambda_j > 0$ and $\sum_{j=1}^C \lambda_j = 1$. The weights in $\pmb{\lambda}$ act as a simple and flexible tuning mechanism that facilitates the trade-off in error protection among message classes of different importance. 

\subsection{Converse \&\ Achievable Finite Blocklength UEP Regions: AWGN Channel}
In \cite{sheldon_2024}, the authors derived and presented both the achievable and the converse finite blocklength (FBL) rate regions for the bit-wise UEP codes over the AWGN channel, and we included them here to establish a rigorous reference point for assessing the performance of our proposed architecture. These results are derived under the assumptions considered herein, including the per-codeword power constraint $\boldsymbol{x} \in \mathcal{X}$ (see Definition 1 and Appendix A in \cite{sheldon_2024}). For simplicity, as in \cite{sheldon_2024}, for the remainder of this paper we focus on the case of $C=2$ message classes, i.e., $\boldsymbol{s}=(\boldsymbol{s}_1, \boldsymbol{s}_2)$ where $|\mathcal{S}_1|=2^{k_1}$ and $|\mathcal{S}_2|=2^{k_2}$ (although the approach can be easily extended to more than two classes).

\begin{figure*}[!t]
    \centering
    \captionsetup{justification=centering}
    \includegraphics[width=0.7\linewidth]{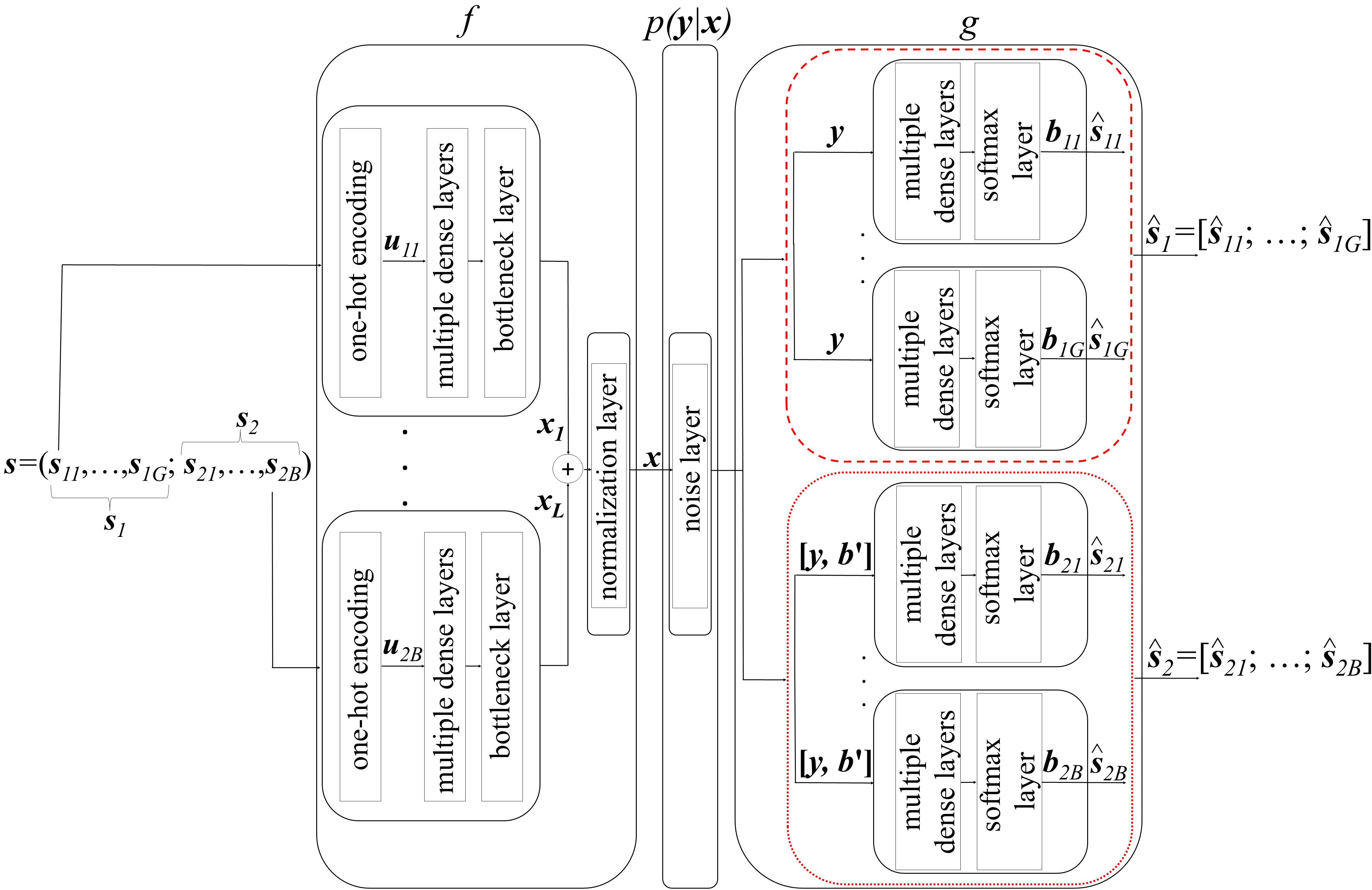}
    \caption{Structured AE-based communication system for bit-wise UEP code design: Important (red dashed) and less important (red dotted) submessage decoding.}
    \label{fig:sys_mod}
\end{figure*}

The regions in \cite{sheldon_2024} are formulated such that, given a triplet $(n, P_{\textrm{e}}^{(1)}, P_{\textrm{e}}^{(2)})$,  they determine the set of achievable rate pairs $(R_1,R_2)$,  along with their sum rate, $R=R_1+R_2$. Conversely, these regions specify which rate pairs are unattainable. To align with our work, we reformulate these results as follows: given a triplet $(n, R_1, R_2)$, determine the achievable and converse bounds on the error probability pairs $(P_{\textrm{e}}^{(1)}, P_{\textrm{e}}^{(2)})$, as well as $P_{\textrm{e}}^{(1)}+ P_{\textrm{e}}^{(2)}$. 

\begin{prop}
    Authors in \cite{sheldon_2024} derived the best--known converse FBL error probability region ($\mathcal{E}^{(C)}$) using EEP codes (also referred to as homogeneous codes in \cite{sheldon_2024}). Specifically, a point-to-point bit-wise UEP code cannot operate at a rate beyond the limits defined by a EEP (homogeneous) code, which is designed to operate with a total error probability equal to the sum of the individual maximum error probabilities of the bit-wise UEP code \cite{sheldon_2024}. For the detailed proof, we refer to \cite{sheldon_2024}, Appendix A, where the $\mathcal{E}^{(C)}$ is derived as:
    \begin{subequations}
    \begin{align}
        \mathcal{E}^{(C)}(n, R_1, R_2)=\bigcup\big\{(P_{\textrm{e}}^{(1)}, P_{\textrm{e}}^{(2)})\in\mathbb{R}^2_+:\nonumber \\  P_{\textrm{e}}^{(1)}+P_{\textrm{e}}^{(2)}\geq Q(\frac{C(\gamma)-R_1-R_2}{\sqrt{\frac{V(\gamma)}{n}}}),\\P_{\textrm{e}}^{(1)}\geq Q(\frac{C(\gamma)-R_1}{\sqrt{\frac{V(\gamma)}{n}}}),\  P_{\textrm{e}}^{(2)}\geq Q(\frac{C(\gamma)-R_2}{\sqrt{\frac{V(\gamma)}{n}}})\big\}
     \end{align}
    \end{subequations}
\end{prop}

\begin{prop} To derive an achievable error probability region, the authors of \cite{sheldon_2024} proposed an approach that combines superposition coding at the transmitter and SIC decoding at the receiver. The SIC decoder first decodes the more important message while treating the less important message as noise. Once the more important submessage is successfully recovered, it is subtracted from the received codeword,  enabling the subsequent decoding of the less important message.  For brevity, we omit the detailed proof; however, a more comprehensive explanation can be found in \cite[Section VI]{tuninetti_2023}. The achievable error probability region $\mathcal{E}^{(S)}$ is obtained by utilizing and simplifying Equation (10) from \cite{sheldon_2024} as follows: 
\begin{subequations}
    \begin{align}
        \mathcal{E}^{(S)}(n, R_1, R_2)=\bigcup_{(\alpha, \beta) \in [0,1]} \big\{(P_{\textrm{e}}^{(1)}, P_{\textrm{e}}^{(2)})\in\mathbb{R}^2_+: \nonumber\\ P_{\textrm{e}}^{(1)}\geq Q(\frac{C(\frac{(1-\alpha)\gamma}{1+\alpha \gamma})-R_1-\beta R_2}{\sqrt{\frac{V'(\alpha \gamma,\gamma)}{n}}})\\ P_{\textrm{e}}^{(2)}\geq Q(\frac{C(\alpha\gamma)-(1-\beta)R_2}{\frac{V(\alpha\gamma)}{n}})\big\},
    \end{align}
\end{subequations}
 where $\alpha$ and $\beta$ represent the power and the rate split, respectively, and $V'(\alpha\gamma,\gamma)=V(\gamma)+V(\alpha\gamma)-2V(\alpha\gamma,\gamma)$. 

\end{prop}

\section{Structured AE--Based UEP Code Design}
\label{Stacked_AE}
A major challenge in modern AE-based code design is the limitation on code blocklength. For example, in \cite{Ninkovic_2021}, we analyzed UEP codes with maximum values of $k=14$ and $n=32$. This limitation arises from the use of one-hot encoding, where the dimension $M$ of the vector $\boldsymbol{u}$ (defined in Section II-A1) grows exponentially with the number of bits $k$ ($M=2^k$). In this paper, we consider a novel UEP code design based on a structured AE architecture consisting of AE-based subblocks of limited size. By organizing these subblocks into a superposition-based encoder and an SIC-based decoder, we effectively extend the blocklength $n$, resolving the complexity issues associated with one--hot encoding.
\vspace{-1mm}
\subsection{Structured AE--Based Blocklength Extension}
The key idea behind the proposed approach is to decompose the encoder/decoder into $L$ AE-based building subblocks, as illustrated in Fig. \ref{fig:sys_mod}. The overall encoder and decoder functions, $(f,g)$, are represented as a set of function pairs $(f_l,g_l), l\in\{1, \dots,L\}$, which are jointly optimized through an end--to--end training process.  To achieve this, several components of the system model need to be redefined and adjusted. 

Consider the binary representation $\boldsymbol{s}\in\mathcal{S}$ of a message $m\in\mathcal{M}$. We assume that $\boldsymbol{s}=(\boldsymbol{s}_1, \boldsymbol{s}_2)$, where $\boldsymbol{s}_1$ (important submessage) and $\boldsymbol{s}_2$ (less important submessage) have lengths of $k_1$ and $k_2$ bits, respectively (Section \ref{bit_wise}). To address the one-hot encoding problem, we partition the input binary sequence into $L$ subsequences of fixed length $K$,  each processed by a corresponding AE encoder subblock. The total binary sequence length is given by $k=L\times K$. Consequently, the submessages are further structured as $\boldsymbol{s}_1=(\boldsymbol{s}_{11}, \dots,\boldsymbol{s}_{1G})$ and $\boldsymbol{s}_2=(\boldsymbol{s}_{21}, \dots,\boldsymbol{s}_{2B})$  with 
$G$ and $B$ denoting the number of subsequences assigned to each submessage, ensuring that $L=G+B$ (Fig. \ref{fig:sys_mod}). The length $K$ is selected as a training hyperparameter to balance model complexity and expressiveness. Our experiments show that subblock size has minimal impact on decoding performance, but strongly influences scalability, as smaller 
$K$ reduces memory usage but may limit learning capacity, whereas larger 
$K$ improves expressiveness at the cost of significantly higher resource demands. Submessage lengths are aligned as integer multiples of $K$, such that $k_1=G\times K$ and $k_2=B\times K$.  

The encoding process follows a structured approach, where each of the $L$ AE encoder building subblocks processes one of the $L$ binary subsequences, as illustrated in Fig. \ref{fig:sys_mod}. Each subsequence is first transformed using one-hot encoding, redefining  the set $\mathcal{U}=\{\boldsymbol{u}_{\boldsymbol{s}_1}, \boldsymbol{u}_{\boldsymbol{s}_2}\}=\{\boldsymbol{u}_{11}, \dots,\boldsymbol{u}_{1G},\boldsymbol{u}_{21},\dots,\boldsymbol{u}_{2B}\}$ (Section II-B1), and then processed by a feedforward neural network. At the output of each encoder, a bottleneck layer of length $n$ is applied. To mimic superposition coding,  the outputs of all $L$ encoder subblocks are summed as $\boldsymbol{x}=\sum_{i=1}^{L}\boldsymbol{x}_i, \boldsymbol{x}_i\in\mathbb{R}^n$, effectively combining the encoded representations into a single transmitted codeword (Fig. \ref{fig:sys_mod}). Finally, the resulting codeword is normalized to ensure the energy constraint is met (Sec. \ref{sec:Model}). 

The codeword $\boldsymbol{x}$ is transmitted through an AWGN channel $\mathcal{W}$,  with noise variance $\sigma^2=(2R\gamma)^{-1}$. At the decoder, the decoding process is organized into $L$ AE decoder building subblocks. Inspired by SIC decoding principles, we adopt \textit{SICNet} \cite{Van_Luong_2022}.  Specifically, we first decode the important submessage, where the first $G$ decoders process the received signal $\boldsymbol{y}$ directly, while the remaining $B$  decoders take as input a concatenation of $\boldsymbol{y}$   and the summed softmax outputs of the first $G$ decoders: $\boldsymbol{b}'=\sum_{i=1}^G\boldsymbol{b}_{1i}$ (Fig. \ref{fig:sys_mod}). Since softmax outputs can be interpreted as probability vectors, their summation produces a new vector $\boldsymbol{b}'$, which amplifies the most confidently decoded submessages while suppressing others. This provides a compact representation of the decoded information, aiding the remaining 
$B$ decoders. Finally, the softmax outputs of all $L$ decoders are collected in $\mathcal{B}=\{\mathbf{b}_{11}, \dots\mathbf{b}_{1G}, \mathbf{b}_{21}, \dots,\mathbf{b}_{2B}\}$.

The training process follows the end--to--end  optimization approach described in Section II-A1, where the encoder-decoder pairs $(f_l, g_l), l\in\{1, \dots,L\}$ are trained using SGD with the Adam optimizer at a learning rate of 0.0006. The key difference lies in the adaptation of the compound loss function from Eq. \ref{eq3} to fit the proposed architecture and the two-class scenario. Specifically, the individual cross-entropy loss (Eq. \ref{eq4}) is computed for each encoder-decoder building subblock, with distinct weight parameters: the first $G$ pairs share a weight $\lambda$ while the remaining $B$ pairs are assigned a weight of $(1-\lambda)$:
\begin{align}\label{Stacked_loss}
    \ell(\mathcal{U}, \mathcal{B})=\lambda\sum_{i=1}^G\ell(\boldsymbol{u}_{1i},\boldsymbol{b}_{1i})+(1-\lambda)\sum_{j=1}^{B}\ell(\boldsymbol{u}_{2j},\boldsymbol{b}_{2j})
\end{align}
All subblocks are trained jointly in an end-to-end fashion, and due to their interaction through the shared channel and global loss function, they exhibit stable and consistent behavior across groups, without significant performance imbalance.

\begin{figure}[!t]
	\begin{tikzpicture}
  	\begin{loglogaxis}[width=1\columnwidth, height=7cm, 
	legend style={at={(0.15,1.14)}, anchor= north,font=\scriptsize, legend style={nodes={scale=0.8, transform shape}}},
   	legend cell align={left},
	legend columns=1,   	 
   	x tick label style={/pgf/number format/.cd,fixed,
   	 precision=1, /tikz/.cd},
   	y tick label style={/pgf/number format/.cd,fixed, precision=1, /tikz/.cd},
   	xlabel={$P_{\textrm{e}}^{(1)}$},
    xlabel style={at={(axis description cs:0.5,-0.08)}, anchor=north},
   	ylabel={$P_{\textrm{e}}^{(2)}$},
   	label style={font=\footnotesize},
   	grid=both,   
   	xmin =0.000000000000000005, xmax = 1,
   	ymin=0.0000000001, ymax=1,
   	line width=0.85pt,
   	tick label style={font=\footnotesize},]
    \addplot[line width=0.5mm, red, mark=none] 
   	table [x={x}, y={y}] {./Results/42_72_21_21/conv}; 
   	\addlegendentry{$\mathcal{E}^{(C)} - (k,n)=(42, 72)$} 
    \addplot[line width=0.5mm, blue, mark=none] 
   	table [x={x}, y={y}] {./Results/42_72_21_21/SUP_RS}; 
   	\addlegendentry{$\mathcal{E}^{(S)}  - (k,n)=(42, 72)$} 
   	\addplot[black, only marks, mark=*] 
   	table [x={x}, y={y}] {./Results/42_72_21_21/AE_Based}; 
   	\addlegendentry{AE--based Code - $ (k,n)=(42, 72)$}
        \addplot[line width=0.5mm,dashed, red, mark=none] 
   	table [x={x}, y={y}] {./Results/56_96_28_28/conv}; 
   	\addlegendentry{$\mathcal{E}^{(C)}  - (k,n)=(56, 96)$} 
    \addplot[line width=0.5mm, dashed, blue, mark=none] 
   	table [x={x}, y={y}] {./Results/56_96_28_28/SUP_RS}; 
   	\addlegendentry{$\mathcal{E}^{(S)}  - (k,n)=(56, 96)$} 
   	\addplot[red, only marks, mark=x,  mark options={line width=1.5pt}] 
   	table [x={x}, y={y}] {./Results/56_96_28_28/AE_Based}; 
   	\addlegendentry{AE--based Code  - $(k,n)=(56, 96)$}
   	\node at (-3.5,-14.4) {\small $\lambda=0.1$};
   	\node at (-18.5,-1.8) {\small $\lambda=0.9$};
  	\end{loglogaxis}
	\end{tikzpicture}
	\vspace*{-5mm}
	\caption{Comparison of converse ($\mathcal{E}^{(C)}$) and achievable ($\mathcal{E}^{(S)}$) error probability regions with AE-based UEP code performance for two configurations: ($k_1=21$, $k_2=21$, $n=72$; solid lines and black circle markers) and ($k_1=28$, $k_2=28$, $n=96$; dashed lines and red $\times$ markers) at $\gamma = 5$~dB.}
	\label{Fig_72_equal}
\end{figure}
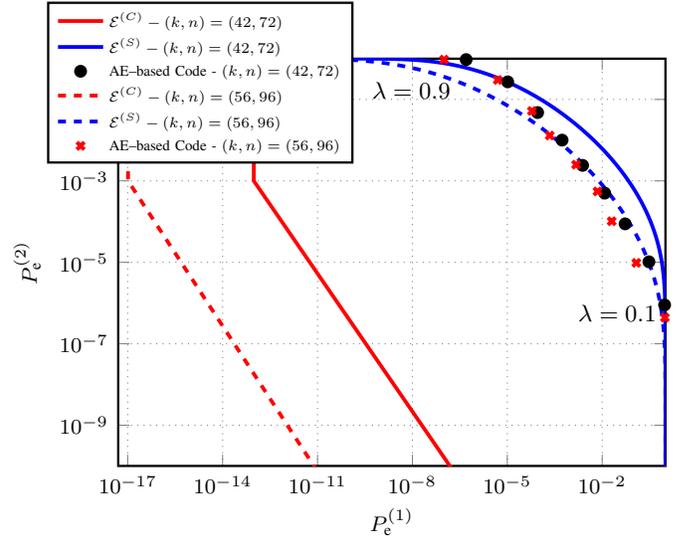

\section{Performance Evaluation}
\subsection{Structured AE--Based Architecture \& Complexity Analysis}
In the architecture described above, each AE encoder subblock processes  $K=7$ bits. Following \cite{OShea_2017}, the training SNR is treated as a network hyperparameter and is set to $\gamma=0$ dB.  As observed in \cite[Section III-B]{Ninkovic_2021}, the choice of $\gamma$ influences the balance between $P_\textrm{e}^{(1)}$ and $P_\textrm{e}^{(2)}$. Each of the $L$ encoder/decoder subblocks contains a single hidden layer with $2^K$ neurons, while the bottleneck layer is of length $n$.  Training is performed on a dataset of $6\times10^5$ samples, and each encoder/decoder subblock is evaluated on $1.5\times10^6$ samples, resulting in a total testing set size of $L\times1.5\times10^6$ samples\footnote{For $\lambda=\{0.1, 0.9\}$, in all figures within Section \ref{Stacked_perf}, the corresponding points were obtained by evaluating the trained models multiple times due to an insufficient number of observed error events.}. The modular structure and limited per-subblock input size allow the architecture to scale efficiently: for instance, at $(k,n)=(56,96)$  (see Section IV-B2), training takes approximately 13 seconds per epoch, the model occupies just 1.77MB of memory, and inference completes in 0.00005 seconds per instance\footnote{Evaluated on a CPU-only laptop (Intel i7-8565U, 16GB RAM).}, confirming its suitability for practical deployment in resource-constrained environments.

\subsection{Structured AE--Based UEP Achievable Rate Region}
\label{Stacked_perf}
 \textit{\indent 1) Code Parameters $(k,n)=(42,72$):} We decompose the encoding and decoding process into $L=6$ subblocks. Fig. \ref{Fig_72_equal} presents  the bit-wise UEP converse error probability region ($\mathcal{E^{(C)}}$, red solid line) along with  the achievable error probability region ($\mathcal{E}^{(S)}$, blue solid line). These regions are computed for $\gamma=5$ dB (following \cite{sheldon_2024}), with submessage parameters $k_1=k_2=k/2=21$ ($B=G=3$). We compare these theoretical error probability regions with the performance of AE-based UEP codes (black circle markers in Fig. \ref{Fig_72_equal}) evaluated in terms of ($P_{\textrm{e}}^{(1)},P_{\textrm{e}}^{(2)}$) pairs at the same $\gamma$, trained using different weight values $\lambda=\{0.1, \dots,0.9\}$ as defined in Eq. \ref{Stacked_loss}. From Fig. \ref{Fig_72_equal}, it is evident that the obtained UEP codes improve the achievable error probability region. Notably, desirable performance characteristics can be easily adjusted by tuning a single parameter $\lambda$ in compound loss function, making the approach simple, efficient and flexible. 

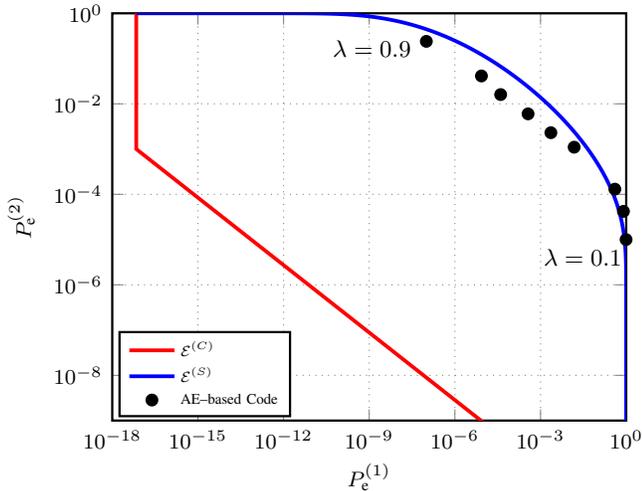
\begin{figure}[t]
	\begin{tikzpicture}
  	\begin{loglogaxis}[width=0.95\columnwidth, height=7cm, 
	legend style={at={(0.18,0.22)}, anchor= north,font=\scriptsize, legend style={nodes={scale=0.8, transform shape}}},
   	legend cell align={left},
	legend columns=1,   	 
   	x tick label style={/pgf/number format/.cd,fixed,
   	 precision=1, /tikz/.cd},
   	y tick label style={/pgf/number format/.cd,fixed, precision=1, /tikz/.cd},
   	xlabel={$P_{\textrm{e}}^{(1)}$},
       xlabel style={at={(axis description cs:0.5,-0.08)}, anchor=north},
   	ylabel={$P_{\textrm{e}}^{(2)}$},
   	label style={font=\footnotesize},
   	grid=both,   
   	xmin =0.000000000000000001, xmax = 1,
   	ymin=0.000000001, ymax=1,
   	line width=0.85pt,
   	tick label style={font=\footnotesize},]
    \addplot[line width=0.5mm, red, mark=none] 
   	table [x={x}, y={y}] {./Results/42_72_14_28/conv}; 
   	\addlegendentry{$\mathcal{E}^{(C)}$} 
    \addplot[line width=0.5mm, blue, mark=none] 
   	table [x={x}, y={y}] {./Results/42_72_14_28/SUP_RS}; 
   	\addlegendentry{$\mathcal{E}^{(S)}$} 
   	\addplot[black, only marks, mark=*] 
   	table [x={x}, y={y}] {./Results/42_72_14_28/AE_Based}; 
   	\addlegendentry{AE--based Code}
   	\node at (-3.5,-12.4) {\small $\lambda=0.1$};
   	\node at (-20.5,-1.8) {\small $\lambda=0.9$};
  	\end{loglogaxis}
	\end{tikzpicture}
	\vspace*{-2.2mm}
	\caption{Comparison of converse ($\mathcal{E^{(C)}}$) and achievable ($\mathcal{E^{(S)}}$) error probability regions with AE--based UEP code performance for ($k_1=14, k_2=28, n=72$) at $\gamma=5$ dB.}
	\label{Fig_72_14}
\end{figure}

To demonstrate the robustness of the proposed approach across different code rates, Fig. \ref{Fig_72_14} presents results for $k_1=14$, $k_2=28$ ($G=2$, $B=4$). Again, the proposed method improves over the known achievable error probability region. Note that for lower weight values ($0.1\leq\lambda\leq0.3$)  improvements are observed primarily in the less important submessage, with minimal impact on the important message. However, as $\lambda$ increases, prioritizing the important message leads to significant gains in both ($P_{\textrm{e}}^{(1)},P_{\textrm{e}}^{(2)}$).

 \textit{2) Code Parameters $(k,n)=(56,96)$:} For the code with $k=56$ bits, encoding and decoding processes consist  of $L=8$ subblocks. Fig. \ref{Fig_72_equal} presents the converse and achievable error probability regions (red and blue dashed lines, respectively), evaluated at $\gamma=5$ dB, with $k_1=k_2=28$ ($B=G=4$). Despite the increased blocklength ($n=96$), the results, shown with red $\times$ markers in Fig. \ref{Fig_72_equal}, remain  comparable to--or slightly improve upon--the achievable error probability region $\mathcal{E}^{(S)}$. However, we observe some degradation for $\lambda>0.7$ particularly in the low-error regime. Evaluating results below $10^{-8}$ can be challenging due to the limited number of error events, leading to potentially less reliable statistical estimation of block error rate.

\section{Conclusion}
This work introduced a structured AE-based bit-wise UEP code design for intermediate blocklengths, whose performance expands the achievable error probability region compared to classical superposition coding-based schemes with SIC decoding. By structuring encoding and decoding into smaller subblocks and incorporating a flexible optimization framework, we provided a scalable approach that effectively balances reliability across different importance classes. While initial experiments suggest that the proposed design can scale to significantly larger blocklengths under limited memory and computational resources, a detailed exploration of this regime is left for future work. However, a significant gap remains between the achievable region and the converse bound. The finite-blocklength UEP capacity remains an open problem, and future work should focus on refining theoretical bounds and optimizing AE-based coding strategies to bridge this gap.

\end{document}